# COVID-19 Detection Using Segmentation, Region Extraction and Classification Pipeline

*Kenan Morani*

*kenan.morani@gmail.com, 0000-0002-4383-5732*

*Electrical and Electronics Engineering Department, Izmir Democracy University, Izmir, Turkey*

**ABSTRACT**

The main purpose of this study is to develop a pipeline for COVID-19 detection from a big and challenging database of Computed Tomography (CT) images. The proposed pipeline includes a segmentation part, a lung extraction part, and a classifier part. Optional slice removal techniques after UNet-based segmentation of slices were also tried. The methodologies tried in the segmentation part are traditional segmentation methods as well as UNet-based methods. In the classification part, a Convolutional Neural Network (CNN) was used to take the final diagnosis decisions. In terms of the results: in the segmentation part, the proposed segmentation methods show high dice scores on a publicly available dataset. In the classification part, the results were compared at slice-level and at patient-level as well. At slice-level, methods were compared and showed high validation accuracy indicating efficiency in predicting 2D slices. At patient level, the proposed methods were also compared in terms of validation accuracy and macro F1 score on the validation set. The dataset used for classification is COV-19CT Database. The method proposed here showed improvement from our precious results on the same dataset. In Conclusion, the improved work in this paper has potential clinical usages for COVID-19 detection and diagnosis via CT images. The code is on github at https://github.com/IDU-CVLab/COV19D_3rd

1. ## INTRODUCTION

COVID-19 affects different people in different ways. Most infected people will develop mild to moderate illness and recover without hospitalization. Most common symptoms of COVID-19 are fever, cough, tiredness, and loss of smell. The virus has been spreading since its onset at the end of 2019. Globally, as of 28 September 2022, there have been 613,410,796 confirmed cases of COVID-19, including 6,518,749 deaths, reported to World Health Organization (WHO). As of September 2022, a total of 12,659,951,094 vaccine doses have been administered [1].

Chest CT scan remains the most sensitive imaging modality in initial diagnosis and management of suspected and confirmed patients with COVID-19. Other diagnostic imaging modalities could add value in evaluating disease progression and monitoring critically ill patients with COVID-19 [2].

In this paper, a pipeline of CT images segmentation, lung area extraction, and classification for COVID-19 detection is proposed.

## 2. RELATED WORK

A lot of research has been done in the area of medical image segmentation and classification. Many researchers used the publicly available dataset 'COVID-19 CT segmentation dataset' [3] for segmentation tasks. Similarly, the private COV19-CT-DB database [4, 5, 6, 7, 8, 9,10] was also used by many researchers for the purpose of COVID-19 diagnosis and detection.

One research work introduced "inf-Net" method as an automated COVID-19 lung infection segmentation including 'COVID-19 CT segmentation dataset'. The proposed neural network aimed to automatically identify infected regions from chest CT scans. The proposed Inf-Net was found to outperform most cutting-edge segmentation models and advances the state-of-the-art performance. The reported Dice score was 0.682 [11].

Another work on the same dataset proposed a method called 'D2A U-Net:' as an automated segmentation solution with dilated Convolution and Dual Attention mechanism. The proposed method outperforms cutting edges methods in semantic segmentation. The method with pretrained encoder achieves a Dice score of 0.7298 and recall score of 0.7071 [12].

Higher Dice score on the same dataset was reported at [13]. The proposed method was a residual attention U-Net for automated multi-class segmentation of COVID-19 infection region. The study provided a promising deep leaning-based segmentation tool to lay a foundation to quantitative diagnosis of COVID-19 lung infection in CT images. The reported dice score was 0.83. Later work improved the dice score to 0.94.

In classification-oriented work on COV19-CT-DB dataset, one paper [14] presented a transformer-based framework for automatic COVID19 diagnosis. The framework in the paper consisted of two main stages: lung segmentation using UNet followed by the classification. The evaluation results show that the method with the backbone of Swin transformer gains the best F1 score of 0.935 on the validation dataset. The final prediction model with Swin transformer achieves the F1 score of 0.84 on the test dataset.

Our work in here proposed a pipeline of segmentation, lung extraction, and classification. Our method better locates lung region of interest using pretrained UNet segmentation model and approach the classification task from the prospective of looking at representative areas of the slices. In that, it combines the work of the authors mentioned above on the same datasets.

## 3. DATASET

The dataset used in this paper is an extension of the COV19-CT-DB database, which includes annotated CT scans of 1,650 COVID and 6,100 Non-COVID cases. The annotation was performed by experts of more than 20 years (4 of them). Each CT scan includes between 50 to 700 slices. In here, we use the original training set and part of the original validation set (partition) of it. The dataset is provided via the "ECCV 2022: 2nd COV19D Competition" [15].

The training set contains, in total, 1992 CT scans and the validation set consists of 494 CT scans. The number of COVID-19 and of Non-COVID-19 cases in each set are shown in Table 1.

The test set contains 4308 CT images, where the labels are not provided.

*Table 1Distribution of cases in training and validation partitions*

| Annotation | Training Data | Validation Data |
|---|---|---|
| *COVID-19 cases* | 882 | 215 |
| *Non-COVID cases* | 1110 | 269 out of 289 |

### 4. METHODOLOGY

The methodology includes a workflow starting with segmentation, lung extraction, and a classification model. A hybrid model was built with one majority voting on three different methodologies and is also tested on the validation set.

The work was entirely implemented on a workstation using GNU/Linux operating system on 64GiB System memory with Intel(R) Xeon(R) W-2223 CPU @ 3.60GHz processor.

*4.1 Segmentation Part[1]*

Traditional segmentation methods as well as UNet architecture-based segmentation were used, and their performances were compared in terms of dice similarity. Region based, thresholding-using-Otsu-method based, and clustering with k=2 based segmentation methods were used. Moreover, UNet model architecture with 3-level depth was used for segmentation. The results of all these methods were evaluated and compared using the publicly available 'COVID-19 CT segmentation dataset' [3]. From the dataset, only the "Image volumes (308 Mb)", and the "Lung masks (1 Mb)" were used. The slices used for evaluation are the "Lung masks" which were sliced in Z-axial direction and the "Image Volumes" sliced in Z-axial direction. The resulting slices were then segmented using one of the proposed segmentation methods.

The training set included all t2 to t8 images and their corresponding masks; i.e. 745 annotated images and 745 annotated masks. Whereas the test set included all t0 and t1 images and corresponding masks; i.e. 84 annotated images and 84 annotated masks.

---

[1] https://github.com/IDU-CVLab/Images_Preprocessing_2nd

The resulting volume slices were compared against the corresponding lung masks in terms of the dice Coefficient value for two classes. Average dice value and minimum dice value over all test lung mask slices were used to evaluate the segmentation performance. The input images were all grayscale, 224x224 sized.

*4.1.1 Traditional Segmentation Methods*

The traditional methods used were region-based, histogram thresholding via otsu, and k-means clustering based with two clusters [16]. Results of the segmentation methods are very similar. That allowed for accurate solutions as well as lightweight solutions with less deep learning usage. So different pipeline parts were modified aiming at accurate more solution (with UNet Deep learning) as well as lightweight solution (with traditional segmentation method, i.e. k-means clustering based). For the UNet model, adding a batch normalization seems to give slightly better performance and therefore the UNet model from this point onward refers to the original UNet structure with addition of batch normalization.

*4.1.2 UNet Based Segmentation*

The UNet segmentation methods included UNet model with and without batch normalization. The UNet model included level depth of three.

The training epoch steps were chosen as in Equation (1):

$$Training\ epoch\ step = \frac{number\ of\ training\ set}{training\ batch\ size} \quad (1)$$

Where $number\ of\ training\ set = 745$ and $training\ batch\ size = 32$

As for the test set epoch, Equation (2) depicts the value:

$$Test\ epoch\ step = \frac{number\ of\ test\ set}{training\ batch\ size} \quad (2)$$

Where $number\ of\ test\ set = 48$ and $test\ batch\ size = 32$

In addition, the learning rate was decreased exponentially over the training epochs as in equation (3):

$$New\ Learning\ Rate = Initial\ Learnin\ Rate \times e^{-epoch} \quad (3)$$

Where $Initial\ Learnin\ Rate = 0.1$.

The UNet model was trained over 20 epochs.

Furthermore, the 'Adam' optimizer and "Binary Cross Entropy" loss function were deployed to train the UNet model for lung segmentation.

*4.2 Lung Extraction Part*

All the images in the COV19-CT-DB dataset were resized to 224x224 from the original size of 512x512, and all images were converted into grayscale. In order for the pixel intensity range in the COV19-CT-DB images to be similar to the pixel intensity in the public dataset used for segmentation model building the sections above, the pixel intensity values were squeezed to values between 0 and 100 using equation (4):

$$COV19 - CT - DB\ image\ pixel\ intenisity = \frac{original\ intenisity\ \times 100.0}{255.0} \quad (4)$$

The above saved UNet models were used for lung-region extraction from the slices in the COV19-CT-DB database. The "predict" function was used on all the slices in the COV19-CT-DB to result in masks. Those resulting masks were overlayed on their original images to extract the lung regions. This overlaying processing resulted after the following processes:

Border clearness (border_clear), binary erosion with "disk (2)", binary closing with "desk (10)", edges detection using roberts filter, and filling the holes using "ndi.binary_fill_holes". Finally, the masks were overlayed on the original images using OpenCV "bitwise_and".

*4.3 Slice Removal*

After UNet segmentation and the lung extraction, slice removal process was deployed to attempt to remove non-representative slices in each CT scan image. A non-representative slice is a slice with little to no-lung areas.

The slice removal techniques counted on the number of non-dark pixels in each image. A threshold of number of non-dark images for each slice is set in order to keep the slice in the CT scan. If any resulting slice has a number of non-dark pixels less than 42x42 or 1764, then the slice is removed from the CT scan .i.e. considered non representative slice.

Different thresholds were tried and visually checking the reduction of number of slices and the type of representative slices were conducted. Table 2 below shows different threshold and number of CT images level empty in every set upon slices removal.

*Table 2 Slice removal thresholds and number of CT scans left empty in every set upon removing slices*

| Threshold | Validation Set | | Training Set | |
| --- | --- | --- | --- | --- |
| | No of CT images empty in COVID Folder | No of CT images empty in non-COVID Folder | No of CT images empty in COVID Folder | No of CT images empty in non-COVID Folder |
| 45x45 | 3 empty | Not-Checked | Not-Checked | Not-Checked |
| 42x42 | 0 | 0 | 8 | 2 empty |
| 40x40 | 0 | 0 | 0 | 0 |

On the test set/partition, a more complex slice removal method was used, aiming to keep at least one slice in each CT scan. We use the initial threshold value of 1764, the first fallback threshold value of 1000, and the second fallback threshold value of 500. If no valid slices are found with the initial threshold value, the code retries with the first fallback threshold value. If no valid slices are found even with the first fallback threshold value, the code retries with the second fallback threshold value. If no valid slices are found even with the second fallback threshold value, we keep the all slices in the folder/CT; i.e. the CT scan does not change.

*4.4 Classification*

The resulted lung extracted images were used as input to a classifier. The classifier is a convolutional neural network model with four similar hidden layer and a fully connected layer [17].

Both vertical and horizontal flipping were used to achieve data augmentation.

Four 2D convolutional layers with 3x3 filter size, padding='same', batch norm, ReLU activation, and 2D max pooling of size 2x2. The filter size sequenced over the four layers as follows; 16, 32, 64, and 128. These layers were followed by a fully connected layer, which includes flatten layer, dense(256), batch norm, ReLU activation, and drop out of 10%. The output is a dense(1) with sigmoid activation function. The output of the final layer is the class probability for predicting a slice as Non-COVID.

To compile the model, binary cross entropy loss, Adam optimizer, and a learning rate scheduler were deployed. The learning rate schedule decreased in an exponential decay during the training, with initial value of 0.1. Table 3 shows classification training hyperparameters.

*Table 3Classification Training Hyperparameters*

| *Batch Size* | 128 |
|---|---|
| *Input Image Size* | 224x224 |
| *Learning Rate Change* | Exponential Decay as in equation (3) |
| *Optimizer* | Adam |
| *Loss Function* | binary cross entropy |

5. **PERFORMANCE EVALUATION**

The matrix used for evaluating the above-mentioned methods were different between segmentation evaluation and classification performance evaluation. For segmentation evaluation, dice similarity score was used. Average dice score and minimum dice score on the test set slices, which are numbered 84 slices, were used [18]. Those scores were evaluated on the test set allocated manually from the publicly available dataset set as mentioned in the segmentation part section above. For the classification part, the macro F1 score was mainly used for performance evaluation [19].

## 6. RESULTS

The results are discussed in three separate sections: the segmentation results, the Lung extraction results, and the classification results. The segmentation part was validated on the publicly available dataset 'COVID-19 CT segmentation dataset', while the lung extraction and classification results were validated on the private database 'COV19-CT-DB'.

*4.1 Segmentation Results*

Table 4 shows the segmentation results on the 'COVID-19 CT segmentation dataset'. The average dice scores can be compared on the test partition images as well as the minimum dice score. The dataset was used as explained in the segmentation section above. The scores are measured on the test set.

*Table 4Traditional segmentation methods and UNet-based segmentation results on 'COVID-19 CT segmentation' test dataset*

| Segmentation Method | Average Dice Score | Minimum Dice Score |
|---|---|---|
| *Region-based* | 0.893 | 0.745 |
| *Histogram-threshold (otsu) based* | 0.893 | 0.751 |
| *k-means clustering (k=2)* | 0.893 | 0.751 |
| *UNet_model (3Layer Depth)* | 0.968 | 0.920 |
| *UNet_model (3Layer Depth & BatchNorm)* | 0.976 | 0.933 |

The results show that the UNet mode with three layers in depth and batch Normalization shows the best performance. However, k-means clustering also gives good performance as a traditional segmentation method.

*4.2 Lung Extraction Results*

Figure 1 shows the resulting lung extracted images. Results are randomly picked from CT scan#50 of the COVID, training set.

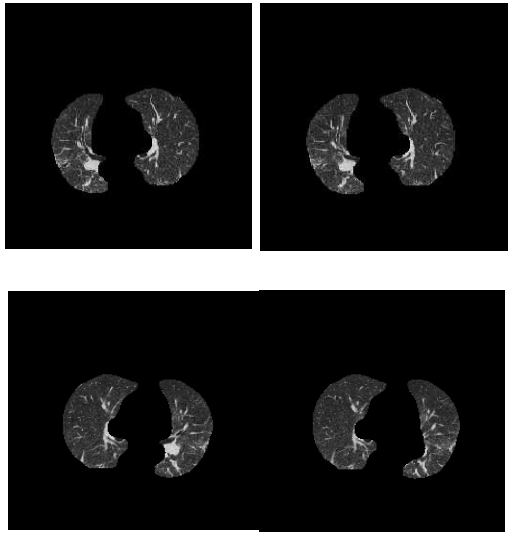

*Figure 1Lung Extracted slices - COVID slices from the training set, using UNet as a segmentation method*

### 4.2 Classification Results

Classification results are discussed at slice-level and at patient level.

#### 4.2.1 Slice-Level Results

Table 5 shows comparison results between different methods for slice classification. Our previous methodology of image processing and CNN modelling [20] gives the highest results at slice level. Input images in all different methodologies are 224x224, except for when slice cropping is used. The input image sizes are indicated in the table.

*Table 5 Comparison of different methods for classification at slice level*

| Method | Method Summary | # of Slices in training/validation sets | Validation Accuracy | Precision | Recall | F1 score | Complexity (# of DL Models) |
|---|---|---|---|---|---|---|---|
| *k-means-cluster-seg-4L-cnn-class* | *Segmentation-Extraction-CNN model; k=2 clustering segmentation, 4-layer depth CNN model* | 436718/106862 | 73.9% | 0.807 | 0.701 | 0.75 | 1x |
| *UNet-seg-4L-cnn-class* | *Segmentation-Extraction-CNN model; Unet-3L-BatchNorm, 4-layer depth CNN model* | 436718/106862 | 77.2% | 0.768 | 0.848 | 0.81 | 2x |
| *UNet-seg-slice-remove-5L-cnn-class* | *Segmentation-Extraction-CNN model; Unet-3L-BatchNorm+slice remove, 5-layer depth CNN model* | 269309/69921 | 83.6% | 0.828 | 0.848 | 0.84 | 2x |
| *Image-process-4L-cnn-class (Our Previous Work 2021)[2]* | *Cropping-Filtering-CNN model* | 436718/106862 | 71.2% | 0.697 | 0.852 | 0.77 | 1x |
| *Image-process-sliceremove-4L-cnn-class (Our Previous Work 2021)* | *Cropping-Filtering-slice removal-CNN model (cropped 255x298 input images)* | 288776/69921 | **85.1%** | 0.839 | 0.910 | 0.87 | 1x |

It is important to note that batch normalization was added to the original UNet structure. It is also important to see that all CNN models contained 4 convolutional layers, except for the case of "UNet-seg-slice-remove-5L-cnn-class", where the CNN model contained 5 convolutional layers.

The threshold for selecting the representative slices in the segmented and lung extracted images "UNet-seg-slice-remove-5L-cnn-class", is taken from the number of non-dark pixels in the images/slices. In that, if the image (224x224) has more than the threshold of (42x42=1764) of non-dark pixels, it is kept in the CT scan (considered representative). Else the slice is removed from the CT scan (non-representative). This threshold resulted from trials and adjusting to make sure that at least one slice is left in most of the CT scans in the remaining and validation sets. Table 6 different thresholds and the resulting number of empty CT scans after thresholding. The empty CT scan folders were deleted before training the model. It is important to note that in this method 5 layers CNN model was used.

---

[2] https://github.com/IDU-CVLab/COV19D

*Table 6 Number of empty CT scans in training and validation sets after different thresholding values when using UNet-based segmentation*

| Threshold | Validation | | Training | |
| --- | --- | --- | --- | --- |
| | No of CT images empty in COVID Folder | No of CT images empty in non-COVID Folder | No of CT images empty in COVID Folder | No of CT images empty in non-COVID Folder |
| 45x45 | 3 empty | Not-Checked | Not-Checked | Not-Checked |
| 42x42 | 0 | 0 | 8 | 2 |
| 40x40 | 0 | 0 | 0 | 0 |

To further validate the results for validation accuracy in our best performing method, the confidence interval for the resulting macro F1 score, as in equation (5), was used for two different number of samples. In the equation, z is taken as z=1.96 for a significance level of 95%. By that we can calculate the confidence interval for the macro F1 score:

*Image-process-sliceremove-4L-cnn-class*

$$interval = 1.96 \times \sqrt{\frac{0.87\,(1 - 0.87)}{69921}} \approx 0.0025$$

*UNet-seg-4L-cnn-class*

$$interval = 1.96 \times \sqrt{\frac{0.81\,(1 - 0.81)}{106862}} \approx 0.06 \qquad (5)$$

The denominator indicates the number of samples in the dataset.

Further confidence intervals can be similarly calculated for the other proposed methods.

The results show sufficient confidence in the resulting validation accuracies.

*4.2.2 Patient-Level Results*

The results show that using a 0.5 threshold for a UNet model-based segmentation model followed by lung extraction and a CNN classifier gives the highest macro F1 score and validation accuracy at patient level. Although "Image-process-sliceremove-4L-cnn-class" seemed to give highest accuracy at slice level, it gives only 70.1% validation accuracy and 0.700 macro F1 score at patient level. These results were at 0.5 class probability threshold. Other class probability thresholds did not improve the performance.

Figure 1 shows the validation accuracy results, whereas figure 2 shows macro f1 scores for some methods.

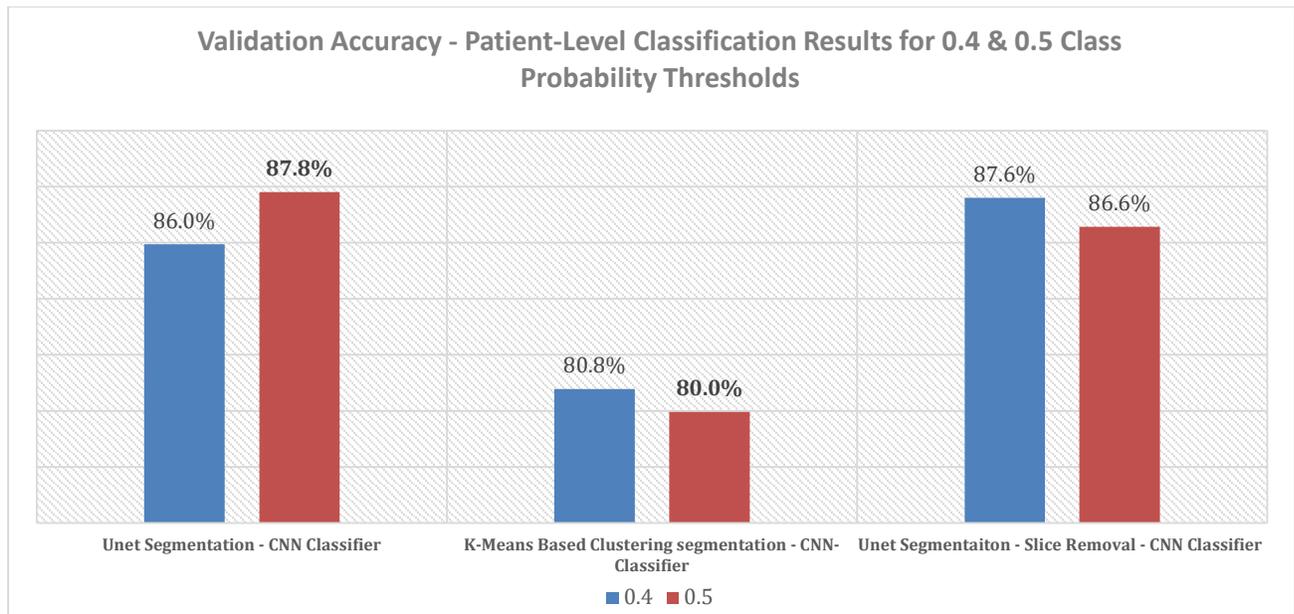

*Figure 2Patient-level classification results in terms of Validation Accuracy the validation set*

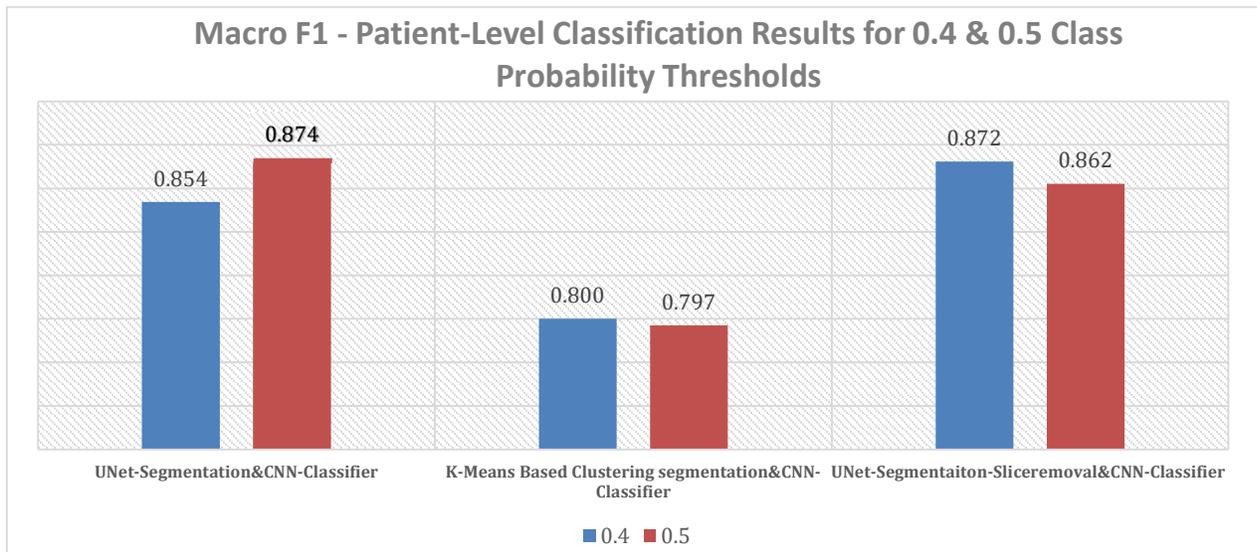

*Figure 3 Patient-level classification results in terms of Macro F1 on the validation set*

The improvement achieved compared to our previous papers [20][21] at patient level is obvious. Table 7 shows the macro F1 score comparison results and the improvement achieved. The comparison was made on the validation set.

*Table 7 Comparing the proposed method (Validation Accuracy/Macro F1) at patient level*

| **Comparing to our previous results** | **Xception Method [21]** | **UNet-Based Segmentation&CNN Classifier without Slice Removal** |
|---|---|---|
| *Macro F1 score on validation* | 0.806 | 0.874 |
| *Validation Accuracy* | 81.10% | 87.81% |
| *Improvement Margin for macro F1* | - | 6.8% |

The results of our methodologies on the test partition in terms of macro f1 score came as in Figure 4. The pipeline methods were tried with 0.4 and 0.5 class probability thresholds [22].

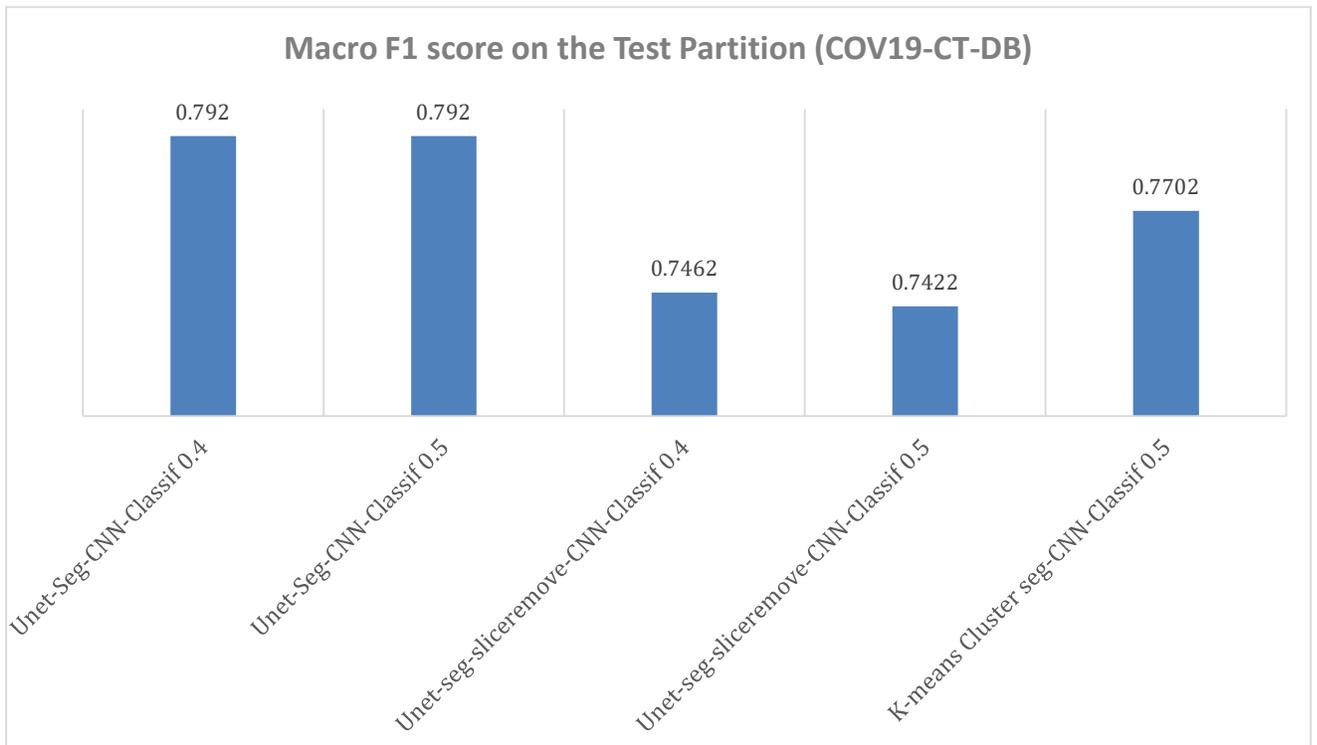

*Figure 4 Macro F1 score on the test partition with 0.4 and 0.5 class probability thresholds*

## 7. CONCLUSION AND FURTHER WORK

In this work, we built a full pipeline for COVID-19 detection via CT images. To achieve that and test the efficiency, different segmentation methods were compared in terms of performance and complexity. The resulting images were used as input for a CNN classifier to take diagnosis decision and classification results were compared in terms at slice-level and at patient-level.

We also introduce a hybrid method using majority voting from our previous and newly reported methods. The resulting hybrid method is more accurate.

Further work includes expanding each methodology proposed in this paper to reach better performance with focus on different goals; higher performance, less complex methodology; a trade-off between high performance and less-complex solutions.

### ACKNOWLEDGEMENT

Acknowledgement goes to the medical staff who worked on annotating COV19-CT-DB database and other members who shared the dataset.

**REFERENCES:**

1. https://covid19.who.int/ {Last Access 28.09.2022}
2. Aljondi, Rowa, and Salem Alghamdi. "Diagnostic Value of Imaging Modalities for COVID-19: Scoping Review." Journal of medical Internet research vol. 22,8 e19673. 19 Aug. 2020, doi:10.2196/19673
3. http://medicalsegmentation.com/covid19
4. Kollias, D., Arsenos, A. and Kollias, S., 2022. Ai-mia: Covid-19 detection & severity analysis through medical imaging. arXiv preprint arXiv:2206.04732.
5. Kollias, D., Arsenos, A., Soukissian, L. and Kollias, S., 2021. Mia-cov19d: Covid-19 detection through 3-d chest ct image analysis. In Proceedings of the IEEE/CVF International Conference on Computer Vision (pp. 537-544).
6. Kollias, D., Bouas, N., Vlaxos, Y., Brillakis, V., Seferis, M., Kollia, I., Sukissian, L., Wingate, J. and Kollias, S., 2020. Deep transparent prediction through latent representation analysis. arXiv preprint arXiv:2009.07044.
7. Kollias, D., Vlaxos, Y., Seferis, M., Kollia, I., Sukissian, L., Wingate, J. and Kollias, S., 2020, September. Transparent adaptation in deep medical image diagnosis. In International Workshop on the Foundations of Trustworthy AI Integrating Learning, Optimization and Reasoning (pp. 251-267). Springer, Cham.
8. Kollias, D., Tagaris, A., Stafylopatis, A., Kollias, S. and Tagaris, G., 2018. Deep neural architectures for prediction in healthcare. Complex Intell Syst 4: 119–131.
9. Arsenos, A., Kollias, D. and Kollias, S., 2022, June. A Large Imaging Database and Novel Deep Neural Architecture for Covid-19 Diagnosis. In 2022 IEEE 14th Image, Video, and Multidimensional Signal Processing Workshop (IVMSP) (pp. 1-5). IEEE.
10. Kollias, D., Arsenos, A. and Kollias, S., 2023, February. AI-MIA: Covid-19 detection and severity analysis through medical imaging. In Computer Vision–ECCV 2022 Workshops: Tel Aviv, Israel, October 23–27, 2022, Proceedings, Part VII (pp. 677-690). Cham: Springer Nature Switzerland.
11. Fan, Deng-Ping, et al. "Inf-net: Automatic covid-19 lung infection segmentation from ct images." IEEE Transactions on Medical Imaging 39.8 (2020): 2626-2637.
12. Zhao, Xiangyu, et al. "D2a u-net: Automatic segmentation of covid-19 lesions from ct slices with dilated convolution and dual attention mechanism." arXiv preprint arXiv:2102.05210 (2021).
13. Chen, Xiaocong, Lina Yao, and Yu Zhang. "Residual attention u-net for automated multi-class segmentation of covid-19 chest ct images." arXiv preprint arXiv:2004.05645 (2020).
14. Zhang, Lei, and Yan Wen. "A transformer-based framework for automatic COVID19 diagnosis in chest CTs." Proceedings of the IEEE/CVF International Conference on Computer Vision. 2021.
15. https://mlearn.lincoln.ac.uk/eccv-2022-ai-mia/
16. Pham, Dzung L., Chenyang Xu, and Jerry L. Prince. "A survey of current methods in medical image segmentation." *Annual review of biomedical engineering* 2.3 (2000): 315-337.

17. Morani, Kenan, and Devrim Unay. "Deep Learning Based Automated COVID-19 Classification from Computed Tomography Images." arXiv preprint arXiv:2111.11191 (2021).

18. Zou, Kelly H et al. "Statistical validation of image segmentation quality based on a spatial overlap index." Academic radiology vol. 11,2 (2004): 178-89. doi:10.1016/s1076-6332(03)00671-8

19. Opitz, Juri, and Sebastian Burst. "Macro f1 and macro f1." arXiv preprint arXiv:1911.03347 (2019).

20. Morani, Kenan, and Devrim Unay. "Deep Learning Based Automated COVID-19 Classification from Computed Tomography Images." arXiv preprint arXiv:2111.11191 (2021).

21. Morani, Kenan et al. "COVID-19 Detection Using Transfer Learning Approach from Computed Tomography Images." (2022).

22. https://drive.google.com/file/d/1ATt-sqsSSaQczz-Qxj85LohwPD3T0i3W/view {Last Access 29.03.2023}